\documentclass[prl,aps,showpacs,twocolumn]{revtex4}
\usepackage{graphicx}
\usepackage{amsmath,amssymb,revsymb}
\usepackage[colorlinks=true,citecolor=blue]{hyperref}

\begin{document}

\title{The Efimov effect for three interacting bosonic dipoles}

\author{Yujun Wang}
\author{J. P. D'Incao}
\author{Chris H. Greene}
\affiliation{ Department of Physics and JILA, University of Colorado, Boulder, Colorado, 80309-0440, USA}

\begin{abstract}
Three oriented bosonic dipoles 
are treated
using the hyperspherical adiabatic representation, providing numerical evidence that 
the Efimov effect persists near a two-dipole resonance and in a system where angular momentum is not conserved. 
Our results further show that the Efimov features in scattering observables
become universal, with a known three-body parameter,  i.e. the resonance energies depend only on the two-body 
physics, which also has implications for the universal spectrum of the four-dipole problem.
Moreover, the Efimov states should be long-lived, which is favorable for their creation
and manipulation in ultracold dipolar gases. Finally, deeply-bound two-dipole states are shown to be relatively stable
against collisions with a third dipole, owing to the emergence of a repulsive 
interaction originating in the 
angular momentum nonconservation for this system.
\end{abstract}
\maketitle

The recent advances in producing ultracold dipolar gases have sparked a great deal of interest in
the novel phases that could be experimentally accessible~\cite{Baranov}. Two key ingredients 
are the long-range and anisotropic nature of the dipolar interaction, which can be manipulated by applying 
an external electric field. This reorients current research scenarios in ultracold gases and opens up the possibility of intriguing phenomena across a broad range of different fields, from condensed 
matter physics to ultracold chemistry. Recent experiments~\cite{DipoleGnd} have been able to create a gas of 
ground-state dipolar molecules and observe the electric field 
and geometry
dependence 
of chemical reactions~\cite{DipoleReact}. Nevertheless, a number of interesting effects can be expected when
the external field is tuned 
near a dipole-dipole resonance (i.e., zero-energy dipolar bound state);
this causes the 
interaction to vary from strongly repulsive to strongly attractive, in close analogy to the 
control through magnetic Feshbach resonances 
\cite{Chin}.

For nonreactive ground-state dipolar molecules~\cite{Stable},  
inelastic three-dipole scattering 
will be the main loss mechanism, and these will determine the stability of the dipolar gas. 
A recent paper by Ticknor and Rittenhouse~\cite{Ticknor2010} predicts in fact that three-body recombination will 
be extremely 
important for bosonic dipolar gases. Three-body recombination in this context, $D+D+D\rightarrow D_{2}+D$, 
involves three free dipoles colliding to form a dipolar dimer releasing enough kinetic energy to make the 
collision partners escape from typical traps. As Ref.~\cite{Ticknor2010} points out, in the strongly
interacting regime universal aspects related to the Efimov physics tend to dictate the collisional aspects of 
such processes. Nevertheless, the persistence of the Efimov effect for long range, anisotropic, dipolar interactions
remains a open question. 
The Efimov effect \cite{Efimov} has proven to have a
profound impact in both nuclear and atomic matter~
\cite{EfimovRefs,Braaten}.
In 
systems with short-range isotropic interactions, the Efimov effect 
manifests when the two-body scattering length $a_{s}$ is much greater than the range of the 
interactions $r_0$, through a series of three-body bound states with energies 
given by 
$E_{n+1}/E_{n}=e^{-{2\pi}/{s_0}}$ ($n=0,1,2,...$),
where the universal constant $s_0$=$1.00624$ for identical bosons.
Note that low-lying energies, however, are not universal and typically depend on the details of the interactions.
In ultracold gases, the Efimov effect is explored near a 
Feshbach resonance
through its impact on 
scattering processes
\cite{Recomb}.

In this Letter, we show that the Efimov effect persists for dipolar systems, and moreover, that the dipolar
interaction is extremely beneficial for the study of Efimov states. 
In particular, we find that the positions of the Efimov resonances (e.g. in energy, in $a_{s}$, etc.), as well as other scattering properties, 
are universally determined by the two-dipole physics alone. Moreover, as the strength of the dipolar interaction 
increases, Efimov states tend to be increasingly long-lived. This scenario, introduced by the dipolar interaction, makes dipolar 
gases ideal systems for the creation and manipulation of Efimov states. Following the method introduced in Ref.~\cite{DIncao}, 
we have also derived the scaling laws for three-body recombination. Finally, we show the existence of an effective 
repulsive long-range interaction between a dipolar dimer and a free dipole, which can 
prevent the decay of such dimers.

\begin{figure*}
\includegraphics[scale=0.7]{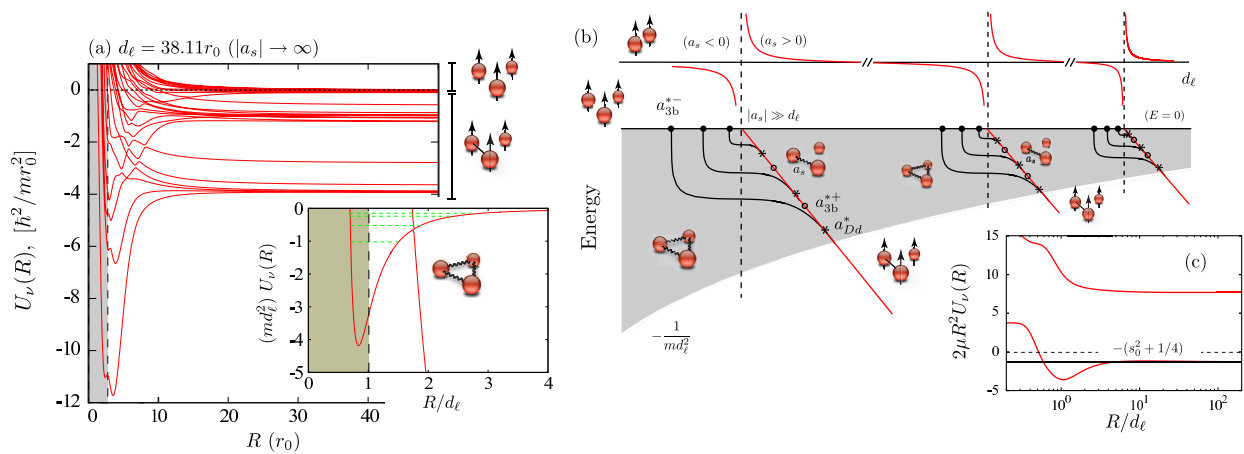}
\caption{(color online) (a) 
A typical set of 
adiabatic hyperspherical potentials $U_\nu(R)$ for three dipoles 
that exhibit the Efimov hyperradial potential curve 
($d_{\ell}/r_{0}=38.11$ and $|a_{s}|\rightarrow\infty$). 
Inset: rescaled adiabatic potentials 
showing the Efimov potential 
(the dashed lines are illustrations of the Efimov states in that potential, not to scale). 
(b) Schematic representation 
for the $d_\ell$ dependence of $a_s$ (upper part) and the Efimov spectrum 
(lower part) 
for dipolar systems (see text). 
(c) The Efimov potential signature at the first pole of $a_s$ ($d_{\ell}/r_{0}=4.86$).  
The horizontal solid line shows 
the Efimov behavior given by Eq.~(\ref{Eq:Efimov}) 
(see text).}
\label{Fig:EfimovPots}
\end{figure*}

The hyperspherical adiabatic representation~\cite{Suno} transforms the 3-body Schr\"odinger equation into
coupled hyperradial 
equations, given (in a.u.) by
\begin{eqnarray}
&&\left[-\frac{1 }{2\mu}\frac{d^2}{dR^2}+U_{\nu}(R)-E\right]F_{\nu}(R)\nonumber\\
&&-\frac{1}{2\mu}\sum_{\nu'} \left[2P_{\nu\nu'}(R)\frac{d}{dR}+Q_{\nu\nu'}(R)\right]F_{\nu'}(R)=0,
\label{SchHyper}
\end{eqnarray}
where the hyperradius $R$ describes the overall size of the system, and $\nu$ 
represents all quantum numbers
that label each channel. In the above equation, $\mu=m/\sqrt{3}$ is the three-body reduced mass, 
$E$ is the total energy, and $F_{\nu}$ is the hyperradial wave function in channel $\nu$. $P_{\nu\nu'}$ and $Q_{\nu\nu'}$ are nonadiabatic couplings that 
drive inelastic transitions, 
while the $U_{\nu}(R)$ are effective potentials that support bound and quasi-bound
states of the system.

The most stringent challenge in this approach is the numerical solution of the hyperangular adiabatic eigenvalue equation at 
fixed values of $R$ in order to determine the potentials 
$U_{\nu}(R)$ and channel eigenfunctions $\Phi_{\nu}(R;\Omega)$,
\begin{align}
\left[\frac{{\hat\Lambda}^2(\Omega)+15/4}{2\mu R^2}+\hat{V}(R,\Omega)-U_{\nu}(R)\right]\Phi_\nu(R;\Omega)=0. 
\label{Adiabatic}
\end{align}
Here $\Omega$ represents the set of all hyperangles describing the 
system's
internal motion, 
$\Lambda^2$ is the grand angular momentum operator~\cite{Suno}
and $\hat{V}=v(\vec{r}_{12})+v(\vec{r}_{31})+v(\vec{r}_{23})$ is the pairwise sum of two-body interactions,
given by
\begin{equation}
v(\vec{r})=V_{0}{\rm sech}(r/r_{0})^2+\frac{2d_{\ell}}{m}\frac{1-3(\hat{z}\cdot\hat{r}) ^2}{r^3} f(r).
\label{TwoDipole}
\end{equation}
The first term 
above
is an isotropic short-range contribution and the second term is the
anisotropic dipole-dipole interaction [with a short-range cut-off $f(r)$] between dipoles 
aligned in the $\hat{z}$ direction with dipole length defined as 
$d_{\ell}=m~{d}_{m}^2/2$,  
where 
${d_{m}}$ 
is the electric dipole moment. 
We emphasize that although our model assumes point dipoles it is also applied
to ground-state dipolar molecules, where all the details of its complicated structure are 
encapsulated in the short-range behavior of the interactions.
Moreover, since the dipoles are free to interact in any geometry, the solutions we obtain
from Eq.~(\ref{Adiabatic}) already include both attractive and repulsive aspects of the dipolar
anisotropic interaction.
To test whether any property of interest is universal with respect to the three-body short-range 
physics, we have performed calculations for different values of $d_{\ell}$ as well as for different
isotropic short-range potentials.

The major difficulty introduced by the dipolar interaction is that the three-body total angular momentum 
$J$ is not conserved.
To calculate $U_\nu(R)$, we expand $\Phi$ as
\begin{equation}
\Phi_\nu^{\Pi,M}(R,\Omega)=\sum_{J,K}\phi_{\nu}^{JK}(R;\theta,\varphi) D_{KM}^J(\alpha,\beta,\gamma),
\label{channelfunc}
\end{equation} 
where $D_{KM}^J$ are the Wigner $D$ functions, $\alpha$, $\beta$ and $\gamma$ are the Euler angles, 
$\theta$ and $\varphi$ are the Smith-Whitten hyperangles~\cite{Johnson}. 
Here $M$ is the space-fixed frame projection of the total angular momentum $J$, 
$K$ is the quantum number for its body-fixed frame projection, and $\Pi$ is the total parity.
Finally, the body frame components $\phi_{\nu}^{JK}$ at fixed $R$ are numerically 
solved by expanding the $\theta$ and $\varphi$ dependences in a B-spline basis.

Here, we study the three-dipole problem for $M^{\Pi}=0^{+}$, which includes the
$J^\Pi=0^+$ symmetry where the Efimov effect occurs for nondipolar systems.
Evidently, for $d_{\ell}\ll r_{0}$ the Efimov effect might be expected to remain unchanged. 
However, 
for $d_{\ell}>r_{0}$ the coupling among different $J$'s might possibly compromise its 
persistence. In order to investigate the Efimov effect for dipoles, we tune $d_{\ell}>r_{0}$ in Eq.~(\ref{TwoDipole}) to get a 
two-dipole zero-energy bound state, 
where $|a_{s}|\gg r_{0}$~\cite{Bohn}. 
Figure~\ref{Fig:EfimovPots} summarizes our findings and sketches the three-body energy spectrum obtained.
Figure~\ref{Fig:EfimovPots}(a) shows the adiabatic potentials $U_\nu$ 
for $d_{\ell}/r_{0}=38.11$ and $|a_{s}|\rightarrow\infty$, showing the overall topology of the potentials, including a family of channels
converging to deeply-bound dimers and channels describing the collision of three free dipoles.
For each dimer channel there exist an infinity of channels converging to that same threshold 
due to the coupling of all different $J$'s 
[we have truncated the $J$-expansion (\ref{channelfunc}) at $J_{\rm max}=14$].

The inset of Fig.~\ref{Fig:EfimovPots}(a), however, shows the evidence of our central result.
It confirms the existence of the universal Efimov potential,
\begin{equation}
U_\nu(R)\simeq-\frac{s_0^2+1/4}{2\mu R^2},
\label{Eq:Efimov}
\end{equation}
which support the usual infinity of three-body Efimov states, illustrated schematically in the figure as horizontal dashed lines. 
Our numerical explorations suggest that, in the strong dipole limit 
where $d_\ell \gg r_0$,  the rescaled potential in the inset of Fig.~\ref{Fig:EfimovPots}(a) is universal, i.e., 
the rescaled potentials for different values of $d_{\ell}$ or for different strengths of the isotropic interaction, 
coincide almost perfectly. 
Figure~\ref{Fig:EfimovPots}(b) 
shows schematically, the three-body energy spectrum versus dipole length. As $d_{\ell}>r_{0}$ increases, new
dimers are created and so are new families of Efimov states. 
The binding energy of the lowest Efimov state scales in proportion to $1/md_{\ell}^2$, based on
our adiabatic potential in the inset of Fig.~\ref{Fig:EfimovPots}(a) as well as Table I. 
Figure～\ref{Fig:EfimovPots}(c) gives further
evidence for the long range universal Efimov potential 
when $|a_{s}|\rightarrow\infty$. 
We checked that adiabatic corrections to the Efimov potential are negligible.

The existence of the Efimov effect for three dipoles might have been expected as a consequence of the 
$s$-wave dominance of the two-body physics near the pole of $a_{s}$~\cite{Ticknor2010}. 
Nevertheless, the Efimov effect for dipoles has a striking difference from the usual Efimov effect. 
As shown in the inset of Fig.~\ref{Fig:EfimovPots}(a) [see also Fig.~\ref{Fig:EfimovPots}(c)], the Efimov 
potential extends {\em only} from values of $R>d_{\ell}$ while for $R<d_{\ell}$ the dipolar interaction 
causes the Efimov channel to be repulsive. 
This critically affects the universal properties 
associated with the dipolar Efimov effect. Since the Efimov states are well separated from the short-range region 
(at $R\approx r_{0}$) by the repulsive barrier at $R<d_{\ell}$, properties of Efimov states are expected to be {\it fully}
universal, i.e., they will depend on the two-dipole physics alone, namely, $a_{s}$ and $d_{\ell}$. 
This implies that, in contrast to the usual Efimov effect, the three-body phase (or parameter)~\cite{Braaten,DIncao} is now universal 
and one can derive the energies of the Efimov states (resonances) for both $a_{s}>0$ and $a_{s}<0$ ($|a_{s}|>d_{\ell}$). 
Moreover, the barrier for $R<d_{\ell}$ also suppresses decay of the Efimov states and, 
therefore, the Efimov states for large values of $d_{\ell}$ are likely to be more long-lived than in the nondipolar scenarios.

\begin{table}
\begin{ruledtabular}
\begin{tabular}{cccc}
$|a_{s}|\rightarrow\infty$ &$d_{\ell}$ ($r_0$) & $m d_{\ell}^2E_0$ & $m d_{\ell}^2\Gamma$ \\
\hline
   & 14.534 & 3.06$\times$$10^{-2}$ & 5.2$\times$$10^{-3}$ \\
   & 25.498 & 3.03$\times$$10^{-2}$ & 6.6$\times$$10^{-3}$\\
   & 38.110 & 2.95$\times$$10^{-2}$ & 3.2$\times$$10^{-3}$ \\ [0.05in] 
\hline
& $a_{\rm 3b}^{*-}/d_{\ell}\approx-8.1$ & $a_{\rm 3b}^{*+}/d_{\ell}\approx1.8$ & $a_{Dd}^{*}/d_{\ell}\approx8.6$\\
\end{tabular}
\end{ruledtabular}
\caption{The positions and the widths of the lowest Efimov state for different values of $d_{\ell}$ and 
the universal 
ratios
for the positions of the Efimov features in three-dipole scattering observables.} 
\label{Tab:EfimovEnergies}
\end{table}

To quantify the three-body parameter for dipolar interactions, we calculate the position and the width of the lowest Efimov resonance 
when $|a_{s}|\rightarrow\infty$. To handle the sharp crossings between 
the Efimov channel and other deeply bound channels [see Fig.~\ref{Fig:EfimovPots}(a) and inset], 
we solve Eq.~(\ref{SchHyper}) by the slow variable discretization (SVD) method~\cite{SVD}, 
and extract the position and width of the Efimov resonances.
In Table~\ref{Tab:EfimovEnergies}, we list 
the ground Efimov state energies $E_0$ and the widths $\Gamma$ for a few values of $d_{\ell}$.
As we have expected, the positions of the Efimov resonances in Table~\ref{Tab:EfimovEnergies} exhibit a universal trend. 
For the widths of these resonances, however, we do not observe nor expect a purely universal behavior, since they encapsulate the decay to
deeply-bound states at smaller distances. Nevertheless, for increasing $d_{\ell}$, the resonance width is
suppressed as $1/d_{\ell}^2$, indicating increased lifetimes of the Efimov states.
The universal three-body parameter $\kappa_*$=$\sqrt{m E_0}$~\cite{Braaten} is then estimated to be 
$0.17/d_{\ell}$ in the limit of $d_{\ell}\gg r_0$. This knowledge of $\kappa_*$~\cite{Braaten} then allows us to determine the 
universal formulas that predict some important three-body scattering observables,
\begin{align}
&K_3^{(a_{s}>0)}\approx \frac{67.1}{e^{2\eta}}[\sin^2[s_0\ln(\frac{a_s}{d_{\ell}})+2.5]+\sinh^2\eta]\frac{a_s^4}{m}, 
\label{K3pos}\\
&K_3^{(a_{s}<0)}\approx \frac{4590 \sinh(2\eta)}{\sin^2[s_0\ln(\frac{|a_s|}{d_{\ell}})+0.92]+\sinh^2\eta}\frac{a_s^4}{m}, 
\label{K3neg}\\
&a_{Dd}^{(a_{s}>0)}\approx(1.46+2.15\cot[s_0\ln(\frac{a_s}{d_{\ell}})+0.86+i\eta])a_s, 
\label{aad}\\
&V_{\rm rel}^{(a_{s}>0)}\approx\frac{20.3 \sinh(2\eta)}{\sin^2[s_0\ln(\frac{a_s}{d_{\ell}})+0.86]+\sinh^2\eta}\frac{a_s}{m}.
\label{vrel} 
\end{align}
Here $K_{3}$ is the rate coefficient for three-body recombination into weakly ($a_{s}>0$)- and deeply ($a_{s}<0$)- bound 
dipole-dimers.
$a_{Dd}$ is the dipole plus dipole dimer scattering length,
and $V_{\rm rel}$ is the corresponding relaxation rate for such inelastic collision processes,
$D_{2}^*+D\rightarrow D_{2}+D$. 
$\eta$ is related to the probability of decay to deeply bound channels \cite{Braaten}.
Equations~(\ref{K3pos})-(\ref{vrel}) now predict the positions of the main features
of such rates that characterize the Efimov physics. For instance, the minima in $K_{3}$ [Eq.~(\ref{K3pos})]
should occur for values of $a_{s}$ given by $a_{\rm 3b}^{*+}=1.8 e^{n\pi/s_{0}}d_{\ell}$ ($n=0,1,2,...$). 
The ratio $a_{\rm 3b}^{*-}/d_{\ell}$ determines 
the position of the Efimov resonances for $K_{3}$ ($a_{s}<0$) [Eq.~(\ref{K3neg})] while $a_{Dd}^{*}/d_{\ell}$ determines the position
of dipole plus dipole-dimer Efimov resonances [Eqs.~(\ref{aad}) and (\ref{vrel})]. 
The second part
of Table~\ref{Tab:EfimovEnergies} lists these properties in terms of ratios for all processes 
(we have dropped the 
factor $e^{n\pi/s_{0}}$ for simplicity). 
The universality on the three-dipole problem
also has key implications for its four-body analog. 
In particular, the positions of the four-dipole features~\cite{4BUniversal} can be obtained from our results in Table~\ref{Tab:EfimovEnergies}.

\begin{figure}
\includegraphics[scale=0.45]{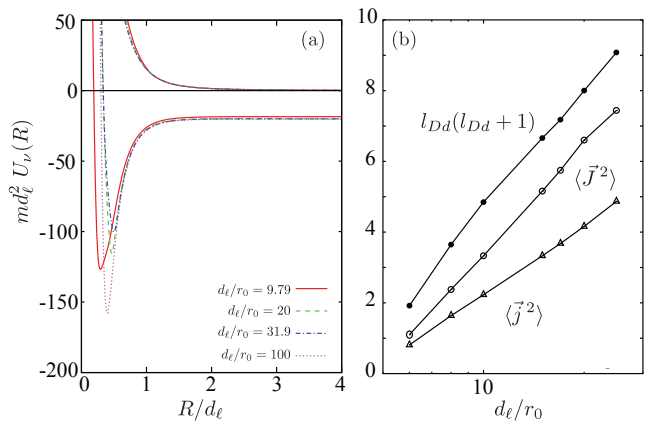}
\caption{
(color online) (a) The diabatized hyperspherical potentials 
for the most weakly-bound dipole plus dipole-dimer channel 
with different values of $d_{\ell}$
and $|a_s|\leq r_0$. (b) The $d_{\ell}$ dependence of the asymptotic angular momentum $l_{Dd}$ 
and the expectation value of the angular momentum $\langle \vec{j}^2 \rangle$ for a deeply-bound dipolar dimer and
the total angular momentum $\langle \vec{J}^2 \rangle$. }
\label{Fig:DipolePots}
\end{figure}

Our numerical results show that the attractive $R^{-2}$ Efimov potential is absent when $|a_s|\lesssim d_{\ell}$. Nevertheless,  
the adiabatic potentials are still universal. 
As shown in Fig.~\ref{Fig:DipolePots}(a), the diabatic potentials show universal scaling with $d_{\ell}$.
Figure~\ref{Fig:DipolePots}(a) shows the lowest three-body continuum channel and 
the lowest dipole plus dipole-dimer channel associated with the most weakly-bound dimer
whose binding energy is proportional to $1/md_{\ell}^2$. 
For this case, recombination to such
dimer states proceeds through an inelastic transition near $R\approx d_{\ell}$, leading to the
$K_{3}\propto d_{\ell}^4$ scaling 
for recombination, consistent with Ref.~\cite{Ticknor2010}.
Since the dipole plus dipole-dimer channel is repulsive for $R<d_{\ell}$, 
we expect that recombination to more deeply-bound states will be substantially less 
important since the pathway to this final state would require tunneling through to small $R<d_{\ell}$.

The behavior of the dipole plus dipole-dimer channels when the dimer is deeply-bound also shows
results interesting for dipole and dipole-dimer mixtures.
These channels should be asymptotically described by
\begin{equation}
W_\nu(R)\underset{R\gg r_{0}}{\simeq} E_{d}+\frac{l_{Dd}(l_{Dd}+1)}{2\mu R^2},
\end{equation}
where $E_{d}$ is the dimer energy and $l_{Dd}$ represents its average or effective orbital angular momentum relative to the free particle. 
For three particles with isotropic short-range interactions, $l_{Dd}$ obeys the usual rules of addition
of angular momentum. For instance, if the two-body state is an $s$-wave state, $l_{Dd}=J$.
For dipolar states, however, our results reveal that is no longer true and we rationalize it by the fact 
that the two-body state is {\em not} a pure angular momentum eigenstate.
In fact, our numerical calculations show that $l_{Dd}$ becomes $d_{\ell}$-dependent.  
Figure~\ref{Fig:DipolePots}(b) shows that $l_{Dd}$ increases with $d_{\ell}$ in the same trend as the expectation 
value of the angular momentum $\langle \vec{j}^2 \rangle$ for the dipolar dimer 
and the expectation value of the total (three-body) angular momentum $\langle \vec{J}^2 \rangle$.
However, the $d_{\ell}$-dependence of $l_{Dd}$ is not universal, as it varies with the short-range physics.
Nevertheless, the non-zero $l_{Dd}$ provides an average centrifugal barrier that suppresses 
the dipole plus dipole-dimer relaxation, $V_{\rm rel}\propto k_{Dd}^{2l_{Dd}}r_{0}^{2l_{Dd}+1}$~\cite{DIncao}, where $k_{Dd}\ll r_{0}^{-1}$ 
is the relative wave number. 
This suppression opens up the possibility of creating stable mixtures of dipoles and two-dipole molecules.
Further, with the tunability of the barrier that can be achieved via field control of $d_{\ell}$, the interaction between a dipole and a two-dipole molecule can be 
controlled, providing an innovation for studies in ultracold quantum gases at both the few-body and many-body level.

In summary, we have characterized the Efimov effect for three bosonic dipoles. A key conclusion is that long-range, anisotropic dipolar interactions are predicted to make 
the Efimov physics more universal than in the traditional systems where this effect 
has been studied and observed. Efimov resonances with dipolar interactions are also predicted to be long-lived, 
which should aid their experimental observation and manipulation. The dipolar interactions introduce a tunable effective 
repulsion between a dipole and a dipolar dimer, which can be used to control the collisions between the molecules and also 
to stabilize the ultracold quantum gas of dipolar dimers.

\begin{acknowledgments}
The authors acknowledge the support of the U.S. National Science Foundation and the AFOSR-MURI.
\end{acknowledgments}

\end{document}